\documentclass[prb,superscriptaddress,showpacs,twocolumn]{revtex4-1}
\makeindex
\usepackage{amsmath}
\usepackage{amsthm}
\usepackage{amssymb}
\usepackage{mathrsfs}
\usepackage{float}
\usepackage[pdftex]{graphicx}
\usepackage[utf8]{inputenc}
\usepackage{xcolor}
\usepackage{mathtools}
\providecommand{\includegraphics}[2][width=\textwidth]{$#2$}

\usepackage{supertabular}
\usepackage{comment}

\newcommand{\braket}[2]{\left<#1 | #2 \right>}

\usepackage{xcolor}
\definecolor{citecol}{rgb}{0.0, 0.6, 0.0}
\definecolor{linkurl}{rgb}{0.2, 0.2, 0.8}
\usepackage[colorlinks=true, urlcolor=linkurl,citecolor=citecol]{hyperref}
\def\doi#1{\href{https://doi.org/#1}{doi:#1}}

\def\RR{\mathbb R}

\def\N{{\mathcal N}}

\def\A{{\cal A}}

\def\HH{H'}

\def\Cdot{\!\cdot\!}
\begin{document}

\title{Fixed-phase approximation and Diffusion Quantum Monte Carlo}
\author{F. Delyon}
\affiliation{LPTMC, UMR 7600 of CNRS, Université P. et M. Curie, Paris, France}
\author{B. Bernu}
\affiliation{LPTMC, UMR 7600 of CNRS, Université P. et M. Curie, Paris, France}

\date{\today}
%\tableofcontents
\pacs{02.70.Ss, 02.70.Rr, 02.70.Tt, 71.10.Ca, 31.15.A}
%02.50.Ey	Stochastic processes
%02.70.Ss	Quantum Monte Carlo methods
%02.70.Rr	General statistical methods
%02.70.Tt		Justifications or modifications of Monte Carlo methods
%71.10.Ca	Electron gas, Fermi gas
%31.15.A-	Ab initio calculations

\begin{abstract}
We present a simple approach to the fixed phase method in Quantum Monte Carlo.
This applies to electrons in molecules and electron gas and is straightforwardly extended to the Schrödinger equation with magnetic field.
\end{abstract}
\maketitle
\section{Introduction}
Quantum Monte Carlo methods have been applied to ground state properties in molecules and electron gas\cite{Kalos,Reynolds,CepAd}.
%Here we consider the Schrödinger equation for an electron gas.

The difficulty to find the ground state of the Hamiltonian comes from the antisymmetry of the wave functions.
The fixed-node approximation is a robust method to obtain an upper bound on the ground state energy. 
Starting with a real trial antisymmetric wave function $\Psi$, this method provides a way to sample the best candidate $\Psi\phi$  where $\phi$
is positive symmetric function. Thus we don't get the actual ground state energy, but in many case this method provides relevant results.

In this case the sign problem is solved by restricting the support of $\phi$ to a region where the trial function $\Psi$ is, say, positive.
There are some attempts\cite{Ortiz} to extend this method to complex trial function (for instance, this is necessary if we consider twisted boundary conditions).
This is the so-called fixed-phase approximation.
In this paper, we propose a new insight into the fixed-phase approximation which extends to the Schrödinger equation with magnetic field.

\section{The fixed-phase approximation}
Let $\Lambda=[0,L]^d$ be a box in $\RR^d$ where $d=2$ or $3$.
Let ${\mathscr H}_a$ be the complex Hilbert space of antisymmetric functions on $\Omega=\Lambda^{N_e}$ and
$H$ be the electronic Hamiltonian for $N_e$ electrons in the box $\Lambda$:
\begin{align}
\label{H}
H=H_0+V
\end{align}
where $H_0=-\Delta=-\frac 12\sum_{i=1}^{N_e}\Delta_i$ is the kinetic part and 
$V$ may contain an external field and the electrostatic interaction.
$V(r_1,\ldots ,r_{N_e})$ is invariant through the  permutation of particles. 
$H$ is an unbounded operator and it is well known that Eq.\ref{H} is not sufficient to define $H$ as a self-adjoint operator specially in a finite box.
In the appendix, we briefly recall how to define a domain D(H) to achieve the definition of $H$. Moreover these domains satisfy
 that if $\Psi\in D(H)$ then $\Psi\phi\in D(H)$ provided that $\phi$ is twice differentiable and satisfies periodic boundary conditions.

%Moreover we suppose that $H$ is bounded from below.
Let $\Psi(R)=\Psi(r_1,\ldots ,r_{N_e})$ be a antisymmetric function, usually called the trial function. 
The fixed phase approximation restricts the domain of $H$ to the functions that can be written as $\Psi\phi$ where $\phi$ is a real non-negative function with periodic boundary conditions (see Appendix \ref{appendix}).

So, we just look at the variational problem:
\begin{align}
\label{min}
E=\min_{\phi(R)\ge 0} \frac {(\Psi\phi,H\Psi\phi)}{ \|\Psi\phi\|^2}
\end{align}
Formally the solution $\phi$ satisfies:
\begin{align}
\label{Hphi}
\Psi^* H\Psi\phi+\Psi H^T{\Psi^*} \phi=2\lambda |\Psi|^2 \phi.
\end{align}
Setting $f=|\Psi|^2\phi$, Eq.\ref{Hphi} rewrites:
\begin{align}
\label{opH1}
\HH f=\frac 12\left({\Psi^*} H\frac 1{{\Psi^*}}+\Psi H\frac 1\Psi\right) f=\lambda f
\end{align}
where we used $H^T=H$. 
$\HH$ is a real operator, thus it may be considered as well as  an operator on the complex symmetric functions.
Let $\mathscr H_s$ be  the Hilbert space on symmetric functions with Hermitian product 
\begin{align}
\label{sp}
{\braket gf}_s=\int \frac { g^*(R)f(R)}{|\Psi(R)|^2}\,dR
\end{align}
Let $D(\HH)\subset \mathscr H_s$ be the subset of functions $f$ with periodic boundary conditions such that $\Delta\frac f{|\Psi|^2}$ is in $\mathscr H_s$.
Then $\HH$ is symmetric on the domain $D(\HH)$, that is,  for any 
$f$ and $g$ in $D(\HH)$:
\begin{align}
{\braket g {\HH f}}_s={\braket{\HH g}f}_s
\end{align}
Using the identity:
\begin{align}
\Psi\Delta\frac{f}{\Psi}=\Delta f-2\nabla\left(f\frac {\nabla \Psi}{\Psi}\right)+\frac {\Delta \Psi}{\Psi}f,
\end{align}
where $\nabla$ is the vector $(\nabla_1,\ldots,\nabla_{N_e})$,
the operator $\HH$ rewrites:
\begin{align}
\label{opH12}
\HH f=-\frac 12\Delta  f +\nabla\Cdot\left( f G\right)+E(R)f
\end{align}
with \begin{align}
G(R)&=\Re\frac {\nabla \Psi(R)}{\Psi(R)},\\
E(R)&=-\frac12\Re\frac {\Delta \Psi(R)}{\Psi(R)}+V(R).
\end{align}
Eq.\ref{opH12} deserves some remarks:

$\bullet$ 
Since $H$ is bounded from below, $\HH$ is also bounded from below and essentially\cite{Sim} self-adjoint on the set $D(\HH)$. Thus the functional calculus applies and in particular  $\exp (-t\HH)$ is
well defined. 

$\bullet$ $\HH$ is symmetric with respect to permutations of particles.
Thus if $f$ is a non-negative eigenvector of $\HH$ then $S(f)$, the symmetrized  of $f$, is also an eigenvector of $\HH$ leading to the symmetric solution $\phi=S(f)/|\Psi|^2$ for the initial problem (Eq.\ref{min}).

$\bullet$ $\HH$ is the generator of a diffusion process with drift $G$ and with sources and sinks $E(R)$.
Thus the ground state of $\HH$ is a positive function and this solution may be simulated by random diffusion processes with branching.
In other word $\HH$ is the generator of a positivity preserving semigroup $\exp (-t\HH)$.
 $\HH$ may be seen as the generator of a semi-group of transformations on the space of probability measures on $\Omega$.

$\bullet$ The invariant probability of the diffusion process with drift $G$ is $|\Psi(R)|^2$. Hence it is just the diffusion process  used in the Quantum Variational Monte Carlo.

$\bullet$ If $P$ is the positive ground state of $\HH$ with eigenvalue $E$ then by direct integration of Eq.\ref{opH12}
\begin{align}\label{Emean}E=\frac{\int E(R)P(R) dR}{\int P(R) dR}\end{align}
%is the solution of Eq.\ref{min}.

$\bullet$ if $\Psi$ is real, then $\HH$ is the the usual operator used in  the Quantum Diffusion Monte Carlo for the fixed node problem. In fact, this the worst case since the singularity of $G$ are of co-dimension one (we assume that $\Psi$ is $C^2$). Let $\N$ be the subset where $\Psi=0$. $\N$ splits the box into at least two connected components and  Eq.\ref{min} will have  at least two independent minima.
% The probability to be on the nodal surface for the diffusion process is zero.
$\N$ is a natural boundary\cite{Brei} of the diffusion process; indeed $|\Psi|^2$ is an eigenvector of the generator of the diffusion with eigenvalue $0$. More simply, the diffusion never reaches the nodal surface $\N$. The unitary operator $U:f \rightarrow f/\psi$, from $\mathscr H_s$ to  $\mathscr H_a$ maps $D(\HH)$ into the regular functions vanishing on $\N$.
Thus, in the real case, $P(R)/\Psi^*$ is the ground state of $H$ with the additional boundary condition $\Psi(R)=0$ on $\N$.

Thus the fixed-phase problem is mapped onto a diffusion model with branching. This model has been widely used\cite{Kalos2,Cep,Cep2001} to compute the energy of the fixed-node model and may be used without change to study the fixed-phase problem. 
As in the fixed-node approximation, we obtain a sampling of the distribution $\Psi_0{\Psi^*} dR$ where $\Psi_0=\phi\Psi$ play the part of an approximate ground state of $H$.
This sampling allows to compute the energy as in Eq.\ref{Emean}. To estimate more general observables, one can straightforwardly adapt the Reptation Quantum Monte Carlo\cite{Pierleoni} algorithms.
However, our approach is simple and avoids the resort to the so called projection method.

%In the next section we briefly recall how to tackle Eq.\ref{opH12}. 
In the next section we briefly describe how to handle the Schrödinger equation with magnetic field. 

\section{The Schrödinger equation with magnetic field}
Let $\A$  be a differentiable vector potential (times the electron charge).
The  Schrödinger operator with magnetic field is given by\cite{mbc}:
\begin{align}
H_\A&=\frac 12(-i\nabla-A)^2+V\\
&=H+\frac i2(\nabla\Cdot  A+A\Cdot \nabla)+\frac{A^2}2\\
\end{align}
As previously, $\nabla$ is the vector $(\nabla_1,\ldots,\nabla_{N_e})$ and $A=\left(\A(r_1),\ldots \A(r_{N_e})\right)$. Thus, for instance, $A^2=\sum_{j=1}^{N_e}\A(r_j)^2$.
Eq.\ref{Hphi} becomes:
\begin{align}
\label{HphiM}
{\Psi^*} H_\A\Psi\phi+\Psi H_\A^T{\Psi^*} \phi=2\lambda \Psi{\Psi^*} \phi
\end{align}
since $H_\A^T=H_{-\A}^{}$, Eq.\ref{HphiM} rewrites:
\begin{align}
\HH_{\A}f=\frac 12\left({\Psi^*} H_\A\frac 1{{\Psi^*}}+\Psi H_{-\A}\frac 1\Psi\right) f
\end{align}
Using the idendity:
\begin{align}
\nonumber
\frac 12{\Psi^*}(\nabla\Cdot\! A+A\Cdot\!\nabla)\frac 1{{\Psi^*}} f%&=2\nabla Af-\frac{\nabla \Psi}{\Psi}Af-\frac{\nabla A\Psi}{\Psi}f\\
=&\nabla \Cdot (Af)-\frac{\nabla {\Psi^*}}{{\Psi^*}}\Cdot Af\\
	&-\frac 12(\nabla \Cdot A) f
\end{align}
we obtain:
\begin{align}
\HH_{\A}&=\HH+i\frac{\nabla \Psi}{2\Psi}\Cdot A-i\frac{\nabla {\Psi^*}}{2{\Psi^*}}\Cdot A+\frac{A^2}2\\
&=\HH -\Im \frac{\nabla \Psi}{\Psi}\Cdot A+\frac{A^2}2
\end{align}
Thus the diffusion process for $\HH_{\A}$ is identical to that of $\HH$, but the branching parts differ:
\begin{align}
	E_\A(R)&=E_{\A=0}(R)-\Im \frac{\nabla \Psi(R)}{\Psi(R)}\Cdot A(R)+\frac{A(R)^2}2
\end{align}

Let us check that this model is invariant thru the gauge transformation
\begin{align}
U:\Psi&\longrightarrow \Psi e^{i\alpha}\\
A&\longrightarrow A+\nabla \alpha.
\end{align}
The drift $G$ is invariant, so we have to check that
\begin{align}
\nonumber
-\frac12&\Re\frac {\Delta \Psi}{\Psi}-\Im \frac{\nabla \Psi}{\Psi}\Cdot A+\frac{A^2}2=-\frac12\Re\frac {\Delta \Psi e^{i\alpha}}{\Psi e^{i\alpha}}\\
\label{Gau} 
	&-(\Im \frac{\nabla \Psi}{\Psi}+\nabla\alpha)\Cdot (A+\nabla\alpha)+\frac{(A+\nabla\alpha)^2}2.
\end{align}
Since 
\begin{align}
-\frac12\Re\frac {\Delta \Psi e^{i\alpha}}{\Psi e^{i\alpha}}=-\frac12\Re\frac {\Delta \Psi}{\Psi}+\frac 12(\nabla \alpha)^2+\Im(\frac {\nabla \Psi}\Psi) \Cdot \nabla \alpha
\end{align}
Eq.\ref{Gau} is satisfied.
%$\HH$ becomes
%\begin{align}
%\HH&\longrightarrow\HH+\frac 12(\nabla \alpha)^2+\Im(\frac {\nabla \Psi}\Psi) \nabla \alpha\\
%%&\HH_{A}+\frac 12(\nabla \alpha)^2+\Im(\frac {\nabla \Psi}\Psi) \nabla \alpha+(\Im \frac{\nabla \Psi}{\Psi}+\nabla\alpha,A-\nabla\alpha)f+\frac{(A-\nabla\alpha)^2}2=\HH_{A}
%%(\Im \frac{\nabla \Psi}{\Psi},A)+\frac{A^2}2&\longrightarrow (\Im \frac{\nabla \Psi}{\Psi}+\nabla\alpha,A-\nabla\alpha)f+\frac{(A-\nabla\alpha)^2}2\\
%%&=(\Im \frac{\nabla \Psi}{\Psi},A)+\frac{A^2}2-\frac 12(\nabla \alpha)^2-\Im(\frac {\nabla \Psi}\Psi) \nabla \alpha
%\end{align}
%and thus the formula for  the new $\HH_{A}$ is:
%\begin{align}
%\HH+\frac 12(\nabla \alpha)^2+\Im(\frac {\nabla \Psi}\Psi) \nabla \alpha+(\Im \frac{\nabla \Psi}{\Psi}+\nabla\alpha,A-\nabla\alpha)f+\frac{(A-\nabla\alpha)^2}2=\HH_{A}
%\end{align}
\section{Conclusion}
We hope that this note clarify the Diffusion Monte Carlo and may be of help to extend and simplify the future numerical implementations.
In particular, Eq.\ref{Emean} gives a straightforward mean to compute the fixed-phase energy avoiding the use of semi-group $e^{-tH}$.

\begin{acknowledgements}
The authors would like to thank M. Holzmann for useful discusions. 
\end{acknowledgements}

\appendix
\section{}
\label{appendix}
%For sake of simplicity, we first suppose that $V$ is bounded (or more generally  $H_0$-relatively bounded\cite{rel})
% so that we may only consider various domains $D(H_0)$ such that the unbounded operator $H_0$ (consequently $H$) is self-adjoint\cite{sa}.
At first, let us recall the domains $D(H_0)$ of self-adjointness of $H_0=-\Delta$.
Firstly, for any $\Psi$ in $D(H_0)$,  $H_0\Psi$ must be in $\mathscr H_a$, that is for each $i$, $\Delta_i \Psi$ is square integrable.
Moreover, $H_0$ must be symmetric, that is for any $\psi_1$, $\psi_2$ in $D(H_0)$ 
\begin{align}
{\braket {\psi_1} {H_0 \psi_2}}={\braket{H_0 \psi_1}{\psi_2}}
\end{align}
which rewrites 
\begin{align}
\label{sym}
\int_{\partial \Omega}d{\bf S} \left(-\psi_1^*\nabla\psi_2+(\nabla\psi_1)^*\psi_2\right)=0
\end{align}
where $\nabla \Psi$ is defined at the 
 boundary as left (or right) derivative. 
 
%Setting  $u_i=\psi_i+i{\bf n}\Cdot \nabla\psi_i$ and  $v_i=\psi_i-i{\bf n}\Cdot \nabla\psi_i$, Eq.\ref{sym} becomes
%\begin{align}
%\label{sym2}
%\int_{\partial \Omega}dS \left(u_1^*u_2-v_1^*v_2\right)=0
%\end{align}
% Eq.\ref{sym2} has many solutions that can be described as the set of unitary operators of $L^2(\partial \Omega)$ mapping $u$ on $v$.
$D(H_0)$ makes $H_0$ self-adjoint if $D(H_0)$ is maximal (any larger subspace will violate Eq.\ref{sym}).
This is the case for following boundary conditions.
 \begin{align}
 \label{twist}
\Psi(r_1,\ldots,Le_j,\ldots ,r_{N_e} )=e^{i\alpha_j}\Psi(r_1,\ldots,0,\ldots ,r_{N_e} )\\
 \label{twist2}
\partial_j \Psi(r_1,\ldots,Le_j,\ldots ,r_{N_e} )=e^{i\alpha_j}\partial_j  \Psi(r_1,\ldots,0,\ldots ,r_{N_e} )
\end{align}
 where $e_j$ is the unit normal vector at $(r_1,\ldots,0\ldots ,r_{N_e} )$ and $\alpha_j$ is real. 
 In this note, ``twisted boundary conditions'' refers to  these condition with some non-zero alpha and 
``periodic boundary conditions'' corresponds to $\alpha_j=0$. 
These are the most commonly used boundary conditions for the Schrödinger operator in Solid State Physics.

However, the Dirichlet boundary conditions, corresponding to $\Psi=0$ on $\partial \Lambda$, are also valid and may be relevant in other contexts.

All these choices for $D(H_0)$ ensure that $H_0$ is a self-adjoint operator and that $\phi\Psi$ is also in $D(H_0)$ if $\phi$ is twice differentiable and satisfies periodic boundary conditions.

Different domains yield to different operators $H_0$ in particular different eigenvalues and different unitary groups $e^{itH_0}$.
For instance, is $\Psi$ is a real function, the operator $\HH$ (Eq.\ref{opH1}) with the domain $D(\HH)$ is conjugate to $H$ with the additional Dirichlet boundary condition on $\N$. 

If $V$ is bounded then any domain of self-adjointness $D(H_0)$ is a domain of self-adjointness of $D(H)$,
but when $V$ is unbounded, it may be technically very painful to define rigorously  $D(H)$. 
However an application of the Kato-Rellich theorem\cite{Sim} ensure that  this is still true if $V\ge 0$ or $V$ contains at most Coulomb singularities.
We always assume that $V$ is of this kind.


\begin{thebibliography}{99}
\bibitem{Kalos}
M. H. Kalos, Energy of a Boson Fluid with Lennard-Jones Potentials, Phys. Hev. A {\bf 2}, 250 (1970); \doi{10.1103/PhysRevA.2.250}
\bibitem{Grimm}
R. Grimm and R. G. Storer. Monte-Carlo solution of Schrödinger’s equation. J. Comput. Phys.,  {\bf 7},134 (1971). \doi{10.1016/0021-9991}
\bibitem{Reynolds}
 J. Reynolds, D. M. Ceperley, B. J. Alder, and W. A. Lester, J. Chem. Phys.  {\bf 77}, 5593 (1982);  \doi{10.1063/1.443766}
\bibitem{CepAd}
 D. M. Ceperley and B. J. Alder, Quantum Monte Carlo for molecules: Green's function and nodal release,  J. Chem. Phys {\bf 81}, 5833 (1984); \doi{10.1063/1.447637}
\bibitem{Ortiz}
 Ortiz, G., Ceperley, D. M. and Martin R. M., New Stochastic Method for Systems with Broken Time-Reversal Symmetry; 2-D Fermions in a Magnetic Field, Phys. Rev. Lett. {\bf 71}, 2777 (1993). \doi{10.1103/PhysRevLett.71.2777}
 \bibitem{Sim}
  M. Reed B.Simon, Functional Analysis, Volume 1 (1981), Academic Press, ISBN: 9780125850506
\bibitem{Brei}
L. Breiman, {\it Probability}, Addison-Wesley Publishing Company, 1968 ISBN 0-201-00646-4
\bibitem{Kalos2}
 M. H. Kalos, Quantum chemistry by random walk, J. Chem. Phys.  {\bf 65}, 4121 (1976); \doi{10.1063/1.432868}
\bibitem{Cep}
 D. M. Ceperley, Fermion nodes, J. Stat. Phys.  {\bf 63} 1237 (1991).  \doi{10.1007/BF01030009}
 \bibitem{Cep2001}
 Lin, C. and Zong, F. H. and Ceperley, D. M. ,Twist-averaged boundary conditions in continuum quantum Monte Carlo algorithms, Phys. Rev. E {\bf 64}, 016702 (2001); \doi{10.1103/PhysRevE.64.016702}
\bibitem{Pierleoni}
C. Pierleoni and D. M. Ceperley, Computational Methods in Coupled Electron–Ion Monte Carlo Simulations, ChemPhysChem 2005, {\bf 6}, 1 \doi{10.1002/cphc.200400587}
\bibitem{mbc}
Eq.\ref{sym} becomes 
\begin{align}
\int_{\partial \Omega}dS \left(-\psi_1^*(\nabla+i\A)\psi_2+((\nabla+i\A)\psi_1)^*\psi_2\right)=0
\end{align}
The Dirichlet boundary conditions are still valid.
Depending on the properties of $\A$, other boundary conditions can be chosen\cite{Ortiz}.
%and valid boundary conditions are obtained by replacing $\nabla$ by $(\nabla+i\A)$ in Eq.\ref{twist2}.
Here again these domains have to be invariant thru the multiplication by a real function $\phi$ with periodic boundary conditions.

\end{thebibliography}
\end{document}